\documentstyle[preprint,revtex]{aps}
\tightenlines
\begin{document}
\noindent
{\it Submitted to:} \hfill {DOE/ER/40322-175}

\noindent
{\it Comments on Nuclear and Particle Physics} \hfill {U. of MD PP \#93-019}

\vspace{1cm}

\begin{title}
{\LARGE The Pion Cloud In Quenched QCD}
\end{title}

\author{\sc Thomas D. Cohen and Derek B. Leinweber}

\begin{instit}
Department of Physics \\
University of Maryland, College Park, MD 20742
\end{instit}

\vspace{36pt}

\centerline{August, 1992}


\begin{abstract}
   Diagrammatic methods and Large $N_c$ QCD are used to argue that the
nucleon as calculated in quenched QCD contains physics which can be
ascribed to the pion cloud of the nucleon.  In particular, it is
argued that the physics corresponding to one-pion loops in chiral
perturbation theory are included in the quenched approximation.
\end{abstract}


\section{Introduction}

   At the present time, the most promising technique for eventually
deriving the low energy properties of hadrons directly from quantum
chromodynamics (QCD) is via numerical Monte Carlo simulations of the
functional integral in a lattice regularized version of the theory.
Unfortunately, given the computational power currently available, it
is not possible to calculate hadron properties in a completely
realistic manner in which quarks of physical mass move in a large
lattice volume with lattice spacings fine enough to guarantee
asymptotic scaling.  A common approximation that significantly reduces
the numerical demands of the simulations is the so-called quenched
approximation.  The computational gains associated with this
approximation allow one to greatly improve on systematic and
statistical uncertainties, and to probe the extremes of the parameter
space.

   In the Euclidean space formulation of the QCD functional integral,
the weighting function has the following continuum form
\begin{equation}
W = \det \left [ \not\!\!D + m_q \right ] \exp (-S_{YM}) ,
\label{fullqcd}
\end{equation}
where $S_{YM}$ is the Euclidean action for a pure Yang-Mills theory
({\it i.e.\ } only gluons).  Integration over the quark Grassmann
fields gives rise to the functional determinant.  The quenched
approximation consists of setting the functional determinant to a
constant, independent of the gluon field configuration.  Thus, in the
quenched approximation, the weighting function is simplified to
\begin{equation}
W_{q} = \exp(- S_{YM}) \, .
\label{quenched}
\end{equation}
This form is obviously amiable to Monte Carlo integration techniques.

   From a diagrammatic point of view, the omission of the fermion
determinant in the weighting function corresponds to the neglect of
all diagrams containing closed quark loops which are not connected to
external sources.

   Given the fact that quenched calculations will continue to be used
for some time to come, it is an important practical problem to
determine what, if any, essential physics is lost when one makes the
quenched approximation.  This issue is particularly important if one
wishes to use the results of quenched lattice QCD calculations to gain
insight into what physics should be included in {\it models} of
hadrons.

   The physics of the pion cloud is believed to play an essential role
in hadronic structure.  Several models of the nucleon, including the
Skyrmion \cite{zahed86}, the cloudy bag model \cite{thomas84}, the
hybrid or chiral bag model \cite{vepstas90} and chiral quark-meson
models \cite{banerjee87} stress the role of the pion cloud in the
nucleon structure.  Thus, the question of whether the quenched
approximation to QCD contains the physics of the pion cloud becomes
important.  In this comment, we will argue that pion cloud effects
are, in fact, included in quenched calculations of nucleon properties.
This result may seem to be counterintuitive, but as shall be shown,
this conclusion is reached both in a large $N_c$ analysis and in an
explicit study of diagrams.  The arguments are somewhat heuristic but
we believe rather compelling.

   Apart from the role of the pion cloud in various models, there is
also a systematic method to estimate the role of pions in hadronic
observables, namely chiral perturbation theory ($\chi$PT).  This
approach is an expansion which is based on a separation of scales
between the pion mass and other scales in hadronic physics
\cite{gasser88,gasser84,weinberg79,langacker73}.  This approach should
rigorously reproduce the low energy properties of QCD if the quark
mass (and hence the pion mass squared) is sufficiently light.  Chiral
perturbation theory predicts that certain quantities are {\it
nonanalytic} in the quark mass as one approaches the chiral limit due
to the effects of the infrared behavior of pion loops.  For our
present purpose, we note that one can use the existence of this
nonanalytic behavior as a signature of pion cloud effects.

   Before proceeding it is worth noting that the the need for quenched
calculations is particularly strong if one wishes to study the role of
the pion cloud.  At present, completely realistic lattice simulations
of pion cloud effects are not possible.  The basic difficulty stems
{}from the use of unphysically large quark masses in the lattice
simulations which leads to pions which are heavy and consequently of
short range.

   A number of challenges are presented as one attempts to decrease
the quark mass.  The long range nature of the pion demands larger
physical lattice volumes which ultimately leads to larger lattices if
one wishes to maintain a reasonable lattice spacing.  Furthermore,
calculations of fermion propagators and Monte-Carlo estimates of the
functional determinant become increasingly difficult as the quark mass
drops and convergence of the algorithms slows.

   For finite computer resources one can always investigate larger
lattice volumes and lighter quark masses in the quenched approximation
than in full QCD.  It is likely that there will be a significant
period of time when quenched calculations are able to probe the
parameter space in areas where pion cloud effects are expected to be
significant whereas full QCD simulations will remain restricted to a
more limited parameter space.  Thus, it is important to establish
whether or not quenched calculations are capable of describing the
physics of the pion cloud.

   In this spirit, it is important to make a more restrictive
definition of the quenched approximation.  Within the usual definition
there remain diagrams which are extremely cumbersome to calculate and
as a result are generally omitted in quenched analyses.  Figure
\ref{genquen} displays two such diagrams with quark loops connected to
external sources.  These quark line disconnected diagrams require
knowledge of the spatial diagonal elements of the inverse fermion
matrix whereas a standard quark propagator requires knowledge of a
single column.  Hence, they are computationally quite difficult and
for the purposes of this investigation we will broaden our definition
of the quenched approximation to exclude diagrams of the type
illustrated in figure \ref{genquen}.

   To make contact with physical observables, it is necessary to
extrapolate from the large values of quark mass currently used in
lattice calculations to their physical masses.  As has been stressed
elsewhere \cite{leinweber92c}, nonanalytic terms in $\chi$PT can lead
to important corrections to this extrapolation in calculations of
charge radii and other observables.  It is clearly of importance to
determine whether these chiral corrections to the extrapolations are
present in the quenched approximation or only in full QCD.

\section{Pion Properties}

   The question of whether the physics of pion loops is present in the
quenched approximation has been discussed previously
\cite{bernard91,sharpe92}.  This discussion has been centered on the
properties of pions and specifically whether the pion mass squared
contains a term proportional to $m_q^2 \ln (m_q)$ which is predicted
in $\chi$PT.  The argument is that a new form of $\chi$PT must be
developed to deal with quenched QCD.  In this quenched version of
$\chi$PT there are no nonanalytic effects arising from the pion cloud
surrounding a pion.  This does not mean that nonanalytic behavior is
absent from mesonic observables.  Nonanalytic chiral behavior can have
its origin in the cloud associated with the isoscalar $\eta'$ meson.
While in nature the $\eta'$ is heavy due to anomalies and topological
effects \cite{coleman85}, in the quenched approximation the $\eta'$ is
degenerate with the pion and $\eta'$ loops can yield nonanalytic
chiral behavior.

   This discussion suggests that quenched QCD calculations of pion
properties will {\it not} correctly reproduce the meson cloud effects
of full QCD.  For example, in full QCD the pion charge radius diverges
logarithmically as the quark mass goes to zero \cite{beg72}; quenched
$\chi$PT predicts it will remain finite in quenched QCD.  The key
point is that the $\eta'$ cloud cannot correctly simulate the role of
the pion cloud for electromagnetic properties since the $\eta'$ is
neutral and does not couple directly to photons.

\section{Nucleon Properties}

\subsection{Diagrammatic considerations}

 We will focus our attention on the properties of the nucleon and will
argue that {\it pion} loop effects do contribute to nucleon properties
even in the quenched approximation.  At first thought this may seem
absurd since pion loops require an intermediate state with a minimum
of one $\overline q q$ pair and the formation of quark-antiquark pairs
is {\it apparently} forbidden by the restrictions imposed by the
quenched approximations.  However, as noted in Ref.
\cite{thomas91,sharpe90,morel87} this restriction is only apparent.
Although the quenched approximation limits quark lines to those which
are connected to external currents, the quark propagators used are
fully relativistic Dirac propagators.  Such propagators contain
``$Z$-graphs'' in which the quark is scattered into a negative energy
state and back.  With the conventional hole interpretation of the
Dirac propagator such processes are the creation and annihilation of
$\overline q q$ pairs.  The restriction imposed by the quenched
approximation is merely that once a virtual pair is created along one
quark line it must be annihilated on the same line.  Thus, it is at
least {\it possible} that quenched QCD may contain $\overline q q$
pairs and such pairs might be pionic in nature.

   The question of whether pion-loop physics is included in quenched
calculations of correlation functions ultimately comes down to the
question of whether intermediate states are reached which have overlap
with physical states containing pions plus other hadrons.  This
suggests that one should study old-fashioned energy denominator type
time-ordered diagrams and ask whether one can reach states which can
be expressed as the product of more than one color singlet operator
operating on the vacuum with at least one of these operators having
the quantum numbers of the pion.  It is clear that this is a necessary
but not sufficient condition to establish the presence of pion loop
physics.  In addition, it must be shown that the singularity structure
corresponds to pion plus hadron states.

   At the hadronic level, some of the pion cloud physics which
accounts for nonanalytic behavior in $\chi$PT can be represented as a
single pion loop.  The one-pion-loop physics is represented by the
hadronic time-ordered diagram in figure 2a.  The corresponding diagram
at the quark level commonly thought to give rise to pionic dressings
but not surviving in the quenched approximation is illustrated in
figure 2b.

   Compare this with the quark level diagram for a nucleon correlation
function in figure 2c which survives in the quenched approximation
\cite{thomas91}.  The intermediate state contains a $\overline q q$
structure (at the top of figure 2c) which has both color-singlet and
color-octet pieces.  The quenched nature of the calculation requires
that the antiquark must be the same flavor as one of the original
quarks in the nucleon interpolating field.  Since there are always two
distinct flavors of quark in a nucleon, the $\overline q q$ structure
can be either isospin zero or one.  The three quark structure below
the $\overline q q$ structure has a piece which is color singlet and
isospin one half.  It is highly plausible that the part of this graph
which consists of a color singlet isospin one $\overline q q$ piece
along with a color singlet isospin one half $qqq$ piece has some
overlap with the physical pion-nucleon scattering state.  Hence it
appears that figure 2c contains, among other things, the essential
physics of figure 2a.  Of course, the diagram in figure 2c is only one
of an infinite class of diagrams which appear to have nonzero overlap
with pion-nucleon scattering states.  One can add to the diagram an
arbitrary number of gluons.

   It is worth noting at this point that a similar diagrammatic
analysis would {\it not} give pionic intermediate states for meson
correlation functions.  For example in figure 3a we show a one loop
hadronic diagram for the $\rho $ meson channel.  The imaginary part of
this graph gives the $\rho $ to two pion decay.  In analogy to figure
2c we construct figure 3b.  This diagram certainly contains two
isovector structures.  However, unlike in the nucleon case in figure
2c, these structures do {\it not} have pion quantum numbers.  They are
color 8 or $\overline 3$ {\it diquarks} with baryon number $\pm 2/3$.
Alternatively, one could arrange the diagram as in figure 3c.  There
is a component in which both structures are color singlets but one
sees that at least one of the structures is isoscalar and cannot
represent a pion.  The isoscalar piece has overlap with the $\eta'$
meson discussed in Ref. \cite{bernard91}.  Thus, this simple
diagrammatic analysis shows that the physics of figure 3a cannot be
reproduced in the quenched approximation.  This result is consistent
with Ref. \cite{sharpe90,sharpe92} and Ref. \cite{bernard91}.

\newpage

\subsection{Large $N_c$ Analysis}

   The preceding diagrammatic analysis is suggestive but not
conclusive since it gives no information about the analytic
properties of the diagram and we do not know whether there is any
spectral strength corresponding to nucleon plus pion states.  The
presence of such spectral strength will lead to nonanalytic behavior
in $m_\pi^2$ around zero.  How can one learn whether there is any
nonanalytic behavior with respect to $m_\pi^2$ in quenched QCD given
the fact that explicit simulations with light quark masses are
currently impractical?  Large $N_c$ QCD \cite{hooft74,witten79}
provides considerable insight.  The key point is that QCD to leading
order in a $1/N_c$ expansion {\it is} quenched.  As shown by `t Hooft
\cite{hooft74}, in any diagram there is a $1/N_c$ suppression factor
associated with each closed fermion loop in a large $N_c$ expansion.
Thus the diagrams which contribute to the leading order expression for
any correlation function have the minimum number of fermion loops.
This is precisely the condition imposed by the quenched approximation.
It is amusing to note that the nonquenched diagram of figure 2b,
commonly thought to give rise to pionic dressings of the nucleon are
$1/N_c$ suppressed relative to that of figure 2c.

   Of course, the leading order large $N_c$ approximation is a more
drastic approximation than the quenched approximation since some
graphs which do not contain internal fermion loops ({\it e.g.} some
non-planar graphs) are also $1/N_c$ suppressed.  Since the large $N_c$
approximation is more severe than the quenched approximation and
contains the quenched approximation it is clear that if there is pion
loop physics (as evidenced by nonanalytic behavior in $m_\pi^2)$ in
large $N_c$ QCD, then the same physics should be present in the less
severe quenched approximation.

   There are two distinct arguments which suggest that pion loop
physics is present in the leading order large $N_c$ approximation.
One way is to study large $N_c$ hadrodynamics
\cite{cohen89,kiritsis89,arnold90} ({\it i.e.\ } a dynamical model based
on hadron degrees of freedom).  The other is via the study of models
such as the Skyrme model which capture the correct leading order $N_c$
physics from QCD.

   The basic idea of large $N_c$ hadrodynamics is that if one produces
an effective hadronic model which reproduces the underlying physics of
QCD then all of the parameters of this hadrodynamical model must scale
with $N_c$ in the manner prescribed by large $N_c$ QCD
\cite{hooft74,witten79}:
\begin{equation}
\Gamma_m^n \sim N_c^{(1-n/2)} ; \quad
M_m \sim 1 ; \quad M_B \sim N_c ; \quad
g_{mBB} \sim N_c^{1/2} ; \quad \Lambda_{Bff} \sim 1 \, ,
\label{hadrodynamics}
\end{equation}
where $\Gamma_m^n$ is a meson $n$-point vertex, $M_m$ a meson mass,
$M_B$ the baryon mass, $g_{mBB}$ a meson baryon coupling and
$\Lambda_{Bff}$ is a baryon form factor mass.  The study of loops in
large $N_c$ hadrodynamics goes back to Witten \cite{witten79} who
showed that meson loops always give corrections to the tree level
meson properties which are suppressed in $1/N_c$.

   Consider the one meson loop contribution to the meson propagator.
The propagators in the loop are all of order unity but from
(\ref{hadrodynamics}) the three-meson vertices are order $N_c^{-1/2}$.
There are two such vertices so the net effect is order $1/N_c$ which
is indeed suppressed compared to the leading order propagator which is
of order unity.  Note however, that this argument does not exclude the
possibility of mesonic loop dressings to mesons in the quenched
approximation since the large $N_c$ approximation is more severe than
the quenched approximation.

   As studied elsewhere \cite{cohen89,kiritsis89,arnold90}, the
situation is more interesting when one studies baryon properties in
large $N_c$ hadrodynamics.  For present purposes, the important issue
is the meson loop contribution to the nucleon mass.  Consider the
nucleon self-energy at the hadronic level.  This receives a
contribution from a one-meson-loop diagram as in figure 2a.  It is
easy to see that this self-energy is order $N_c$.  There are two
$N_c^{1/2}$ factors at the meson-baryon couplings.  The meson
propagator is order unity since the mass is order one and the loop
integral is cutoff by the form factor whose falloff is also order
unity.  In the large $N_c$ limit where nucleon recoil can be
neglected, the nucleon propagator (working in the rest frame of the
nucleon) goes like the inverse of the meson three momentum, {\it i.e.\
} order 1 .  Thus, one sees that in large $N_c$ hadrodynamics (which
is assumed to correctly reflect large $N_c$ QCD), one meson loop
contributions to the nucleon mass are of order $N_c$ which is {\it
leading order} in $N_c$ counting according to (\ref{hadrodynamics}).
As discussed above, the leading order terms in the $1/N_c$ expansion
are quenched and hence the one-meson-loop physics, including
one-pion-loop physics, should be present in the quenched
approximation.

   The preceding argument is consistent with the leading nonanalytic
behavior for pion loops obtained in $\chi$PT.  Perhaps the easiest
place to look for this nonanalyticity is to study $d^2 M_N /
d(m_\pi^2)^2$.  The leading order nonanalyticity from $\chi$PT comes
{}from a one-pion-loop diagram as in figure 2a.  One obtains
\begin{equation}
{d^2 \, M_N \over d(m_\pi^2)^2} =
-{9 \over 128 \pi} \, {g_A^2 \over m_\pi \, f_\pi^2} + {\cal O}(1)
\label{opl}
\end{equation}
It is well known that $g_A \sim N_c$ and $f_\pi \sim N_c^{1/2}$ so
that
\begin{equation}
{d^2 \, M_N \over d(m_\pi^2)^2} \sim N_c
\end{equation}
Integrating twice with respect $m_\pi^2$ gives this pion loop
contribution to $M_N$  scaling like $N_c$ which is the leading
order.

   The preceding argument has a drawback in that it is based on large
$N_c$ hadrodynamics and not directly on QCD.  While it is highly
plausible that the $N_c$ behavior of hadrodynamics (including baryons)
reproduces that of QCD, it has not been proven rigorously.  Ideally,
one should work directly in large $N_c$ QCD which unfortunately is not
possible.  Instead, one can look at models which are believed to
correctly reproduce QCD's large $N_c$ behavior.  The analysis here
will be based on the Skyrme model although all chiral large $N_c$
soliton models of the nucleon ({\it e.g.} the chiral bag model, the
chiral quark meson soliton model) behave the same way.  The issue of
whether the pion cloud physics associated with loops is present in
large $N_c$ QCD or models of large $N_c$ QCD (such as the Skyrme
model) is complicated.  For the present purposes it is reasonable to
assert that pion cloud physics is present in large $N_c$ QCD if
various quantities calculated to leading order in $N_c$ depend on the
pion mass in the same way as pion loop calculations of the same
quantities in a hadronic picture.  The lore of $\chi$PT is that the
leading nonanalytic behavior in $m_\pi^2$ is given by one-pion-loop
graphs in a hadronic model.  Thus, ultimately the issue is whether the
large $N_c$ model calculations produce the same nonanalytic behavior
near the chiral limit as one-pion-loop hadronic calculations.

   Does the Skyrme model correctly reproduce the leading nonanalytic
behavior in $m_\pi^2$ predicted in $\chi$PT?  The answer is a
qualified yes \cite{cohen92b}.  As discussed in detail in Ref.
\cite{cohen92b}, the Skyrme model reproduces the leading nonanalytic
properties of $\chi$PT in the following sense.  If one confines
attention to vector-isovector and scalar-isoscalar operators ({\it
i.e.} whose expectation values which do not vanish in the hedgehog
intrinsic state), and chooses the Skyrme model parameters to give the
correct value of $g_A$, then the Skyrme model prediction for nucleon
matrix elements will precisely reproduce the leading nonanalytic
behavior of $\chi$PT {\it up to an overall factor which depends only
on the quantum numbers of the operator}.

   For scalar-isoscalar operators, the Skyrme model will always give a
coefficient for the leading nonanalytic term which is a factor of
three larger than $\chi$PT.  The origin of this factor of three lies
in the fact that the Skyrme model result depends on the order in which
the chiral and large $N_c$ limits are taken.  In $\chi$PT, it is
explicitly assumed that the pion mass is small compared to {\it all
relevant hadronic scales in the problem}.  Thus, the only physical
states energetically near the nucleon are states with nucleon plus one
pion.  On the other hand, in the large $N_c$ limit for a hedgehog
model the $\Delta$ is degenerate with the nucleon.  The $N$-$\Delta$
splitting goes as $1/N_c$.  Therefore, in the large $N_c$ limit, loops
with $\pi$-$\Delta$ intermediate states should be included along with
$\pi$-$N$ intermediate states when doing $\chi$PT.  The inclusion of
$\pi$-$\Delta$ intermediate states in $\chi$PT, assuming degenerate
$N$ and $\Delta$ masses and using $g_{\pi N \Delta}$ as calculated in
the Skyrme model, precisely accounts for the factor of three
\cite{cohen92b}.  Therefore, at least for this class of observables,
the nonanalytic behavior associated with pion loops is in fact
present.

   An explicit example may help clarify this point.  Once again,
consider $d^2 \, M_N / d(m_\pi^2)^2$.  The one pion loop contribution
in $\chi$PT (including only nucleon intermediate states is given in
(\ref{opl}).  If we also include a diagram analogous to figure 2a with
a $\Delta$-$\pi$ intermediate state and assume $M_\Delta=M_N$ with
$g_{\pi N \Delta}$ given by the Skyrme model value we obtain
\begin{equation}
{d^2 \, M_N \over d(m_\pi^2)^2} =
- {27 \over 128 \pi} \, {g_A^2 \over m_\pi \, f_\pi^2} + {\cal O}(1)
\label{skyropl}
\end{equation}
Here the coefficient is precisely three times the usual $\chi$PT
result.

   One can also calculate $d^2 \, M_N / d(m_\pi^2)^2$ directly from
the Skyrme model.  With the identity
\begin{equation}
{d \, M_{N} \over d(m_\pi^2)} = \bigm < N \bigm | {1 \over 4} \, f_\pi^2
\, tr (U-1) \bigm | N \bigm > \, ,
\end{equation}
along with the asymptotic properties of $U = \exp (i \vec{\tau} \cdot
\vec{\phi} / f_\pi)$ where $\vec{\phi}$ is the nonlinear realization
of the pion field, and the standard Skyrme model expression for $g_A$,
one can compute $d^2 \, M_N / d(m_\pi^2)^2$ rather easily.  The result
is found to be the expression of (\ref{skyropl}).  Thus, we see
explicitly that a $1/N_c$ model such as the Skyrme model does have
nonanalytic behavior in $1/m_\pi^2$ precisely as one anticipates from
a calculation of pion dressings of the nucleon with a $\Delta$
degenerate in mass.

\section{ Summary}

   In summary, we have given heuristic---but compelling---arguments as
to why simulations of nucleons in quenched QCD contain pion cloud
physics which chiral perturbation theory ascribes to one pion loop
contributions at the hadronic level.  We have shown that diagrams
surviving in the quenched approximation contain pieces which ``look
like'' nucleon plus pion states.  We have also argued that one can use
the $1/N_c$ approximation as a surrogate for the quenched
approximation since the $1/N_c$ approximation {\it is} quenched.  Two
distinct arguments, one based on large $N_c$ hadrodynamics and one
based on $1/N_c$ hedgehog models such the Skyrme model, both suggest
that pion loop physics is present in large $N_c$ QCD and hence in
quenched QCD.  These results contrast the superficial perception that
the quenched approximation is incapable of including the physics of
pionic dressings of baryons.

   We thank Wojciech Broniowski and Manoj Banerjee for helpful
conversations.  D.B.L. thanks Richard Woloshyn and Terry Draper for
early discussions which stimulated his interest in these issues.  This
work is supported in part by the U.S. Department of Energy under grant
DE-FG05-87ER-40322.  T.D.C. acknowledges additional financial support
{}from the National Science Foundation though grant PHY-9058487.
\vspace{1in}

\noindent
\hspace*{2.5in}{\it Thomas D. Cohen and Derek B. Leinweber \\}
\hspace*{2.5in}{\it Department of Physics, University of Maryland \\ }
\hspace*{2.5in}{\it College Park , MD 20742}

\newpage

\figure{Skeleton diagrams of disconnected quark loops connected to
external sources which are generally not included in quenched QCD
analyses.  Diagram (a) contributes to the three-point correlation
function of a baryon current matrix element.  Diagram (b) contributes
to the two-point correlation function of an iso-scalar meson.  The
diagrams may be dressed with an arbitrary number of gluons.
\label{genquen}}

\figure{Time ordered diagrams for one-pion-loop dressings of the
nucleon.  Time flows from left to right.  For illustrative purposes we
have selected the proton with a $\pi^+$-$n$ intermediate state for all
three diagrams.  Figure (a) illustrates the hadronic level dressing
and figure (b) describes a quark level diagram na\"{\i}vely thought to
exclusively account for pionic dressings of the nucleon.  This diagram
is excluded in the quenched approximation.  Figure (c) illustrates a
quark level diagram whose quantum numbers overlap with that of figure
(a) and which survives in the quenched approximation.
\label{baryon}}

\figure{Time ordered diagrams for two pion intermediate states of the
$\rho$-meson.  Figure (a) illustrates the hadronic level intermediate
state $\pi^0 \, \pi^+$.  Figure (b) describes a quark level diagram
analogous to that in figure 1c but which has no overlap with that of
figure 2a.  Figure (c) illustrates a quark level diagram which has
overlap with a two meson intermediate state.  However, one of the
mesons is an isoscalar and cannot be identified with a pion.
\label{meson}}

\end{document}